\documentclass[12pt,a4paper]{article}
\usepackage[text={16.5cm,23cm},centering]{geometry}
\usepackage{amsmath}
\usepackage{amsthm}
\usepackage[utf8]{inputenc}
\usepackage{tgtermes,tgheros,tgcursor}
\usepackage[varg]{newtxmath}
\usepackage{hyperref}
\hypersetup{%
 colorlinks=true,%
 linkcolor=blue,
 citecolor=blue,
}
\usepackage{xspace}

\numberwithin{equation}{section}

\newcommand{\Zb}{\mathbb{Z}}
\newcommand{\Rb}{\mathbb{R}}

\DeclareMathOperator{\sign}{\mathrm{sign}}

\newcommand{\omegal}{\omega_{\mathrm{local}}}
\newcommand{\psil}{\chi}
\newcommand{\psibulk}{\phi}

\begin{document}
\begin{center}
  %% Preprint number
  \begin{flushright}
    OU-HET 1259
  \end{flushright}
  \vspace{8ex}
  %% Title
  {\Large \bfseries \boldmath Exotic massive fermionic systems with huge vacuum degeneracy at boundaries}\\
  \vspace{4ex}
  % Author
  {\Large Hiroki Kawakami and Satoshi Yamaguchi}\\
  \vspace{2ex}
  {\itshape Department of Physics, Graduate School of Science, %affiliation
  \\
  Osaka University, Toyonaka, Osaka 560-0043, Japan}\\
  \vspace{1ex}
  \begin{abstract}
    We investigate a massive non-relativistic fermionic system exhibiting exotic features. When the mass parameter is set to zero, the system acquires the fermionic subsystem symmetry.  Introducing the mass term explicitly breaks this symmetry, resulting in a trivially gapped system in the absence of boundaries.  We demonstrate that with the introduction of boundaries, the system remains gapped, but it has a huge vacuum degeneracy.  The residual entropy is proportional to the area of the boundary. 
  \end{abstract}
\end{center}

\vspace{4ex}
\section{Introduction}
Recently, there has been increasing interest in the study of non-relativistic systems with exotic features, including fractons \cite{Chamon:2004lew,Haah:2011drr}.
For nice reviews on the topic, see \cite{Nandkishore:2018sel,Pretko:2020cko,Gromov:2022cxa}.
Not only bosonic systems, fermionic systems are also studied.
Notably, a fermionic model with fermionic subsystem symmetry was introduced in \cite{Yamaguchi:2021qrx} by considering the supersymmetric version of the bosonic models studied in \cite{Seiberg:2019vrp,Seiberg:2020bhn,Seiberg:2020wsg,Gorantla:2020xap}.
Additional studies of similar fermionic models can be found in \cite{Katsura:2022xkg,Honda:2022shd}.
Furthermore, lattice fermionic fracton systems have been explored in \cite{you2019building,Tantivasadakarn:2020lhq,Shirley:2020ass,Cao:2022lig}.

Introducing boundaries to such exotic systems presents an intriguing avenue of study.
For example, the anomaly inflow mechanism \cite{Callan:1984sa} associated with subsystem symmetry gives rise to robust, gapless boundary degrees of freedom in the systems that are otherwise trivially gapped in the absence of boundaries.
These systems, commonly referred to as subsystem symmetry protected topological (SSPT) phases, have been extensively studied in works such as \cite{You:2018oai,Devakul:2018fhz,Devakul:2019duj,Burnell:2021reh,Yamaguchi:2021xeq,Shimamura:2024kwf,Okuda:2024azp,Ebisu:2024eew}.

In this paper, we investigate a different type of boundary degrees of freedom in a system that is trivially gapped in the bulk.
Specifically, we study the $\psi$ theory, an exotic fermionic system introduced in \cite{Yamaguchi:2021qrx}.
By adding a mass term to the $\psi$ theory, the fermionic subsystem symmetry is explicitly broken, resulting in a system that is trivially gapped in the absence of boundaries.
This behavior is analogous to the $(1+1)$-dimensional Majorana fermion \cite{Kitaev:2000nmw}, which is also trivially gapped in a space without boundaries when a mass term is present.
For the Majorana fermion, the presence of boundaries may lead to robust ground state degeneracy depending on the mass term and the boundary condition.
We examine a similar phenomenon in the massive $\psi$ theory.

We observe that when boundaries are introduced to the massive $\psi$ theory, the system remains gapped; however, it exhibits a huge number of ground states. The residual entropy, defined as $\log$(the number of ground states), is proportional to the area of the boundary.  Our findings differ from those in previous works \cite{Burnell:2021reh,Yamaguchi:2021xeq} in the following ways:
\begin{itemize}
	\item The system remains gapped even with the introduction of boundaries.  
	\item The (fermionic) subsystem symmetry is explicitly broken by the mass term.  Consequently, the boundary ground state degeneracy is not protected by this symmetry. If the degeneracy is robust, it must be protected by another, as-yet unidentified, symmetry.
\end{itemize}
Investigating these observations within the framework of SSPT phases is an interesting future work. 

The rest of this paper is organized as follows. In Sec.~\ref{sec:bulk}, we introduce the massive $\psi$ theory and demonstrate that it is trivially gapped in the absence of boundaries.
In Sec.~\ref{sec:boundary}, we examine the effects of introducing boundaries, identify possible boundary conditions, and the spectrum of the system.
Sec.~\ref{sec:conclusion} is devoted to conclusions and discussions.

\section{The bulk system}
\label{sec:bulk}

In this section, we set up the model of the bulk system.  

The system is defined on a $(3+1)$-dimensional spacetime. The three-dimensional space is a torus $T^3$, whose coordinates $x,y,z$ satisfy the following identifications:
\begin{equation}
  (x,y,z) \sim (x+l_x,y,z) \sim (x,y+l_y,z) \sim (x,y,z+l_z),
\end{equation}
where $l_x, l_y, l_z > 0$ are the periods in the $x, y, z$ directions, respectively. The temporal coordinate is denoted by $t$. 

The theory considered in this paper is a massive fermionic theory, formulated as follows.
The fundamental fields of the theory are two real fermionic fields $\psi_{\pm}(t,x,y,z)$. The action is given by
\begin{equation}
	S=\int d^3xdt( i\psi_+\partial_- \psi_+ + i\psi_-\partial_+\psi_- - im\psi_-\psi_+),
	\label{eq:action}
\end{equation}
where $m \in \Rb$ is the mass parameter. $\partial_{\pm}$ are defined as:
\begin{equation}
	\partial_\pm := \frac{1}{2}(\partial_t\pm\partial_x\partial_y\partial_z).
\end{equation}

The theory described by \eqref{eq:action} with $m=0$ has been first introduced in \cite{Yamaguchi:2021qrx} and referred as the $\psi$ theory. In this case, the theory has so called the fermionic subsystem symmetry, which is a shift of $\psi_{\pm}$ by a function of any two spatial coordinates among $x$, $y$, and $z$.
This $\psi$ theory is gapless or equivalently ``massless.'' If periodic boundary conditions are imposed on $\psi_{\pm}$, the system has a huge vacuum degeneracy as discussed in \cite{Yamaguchi:2021qrx}.
In this paper, for simplicity, we impose anti-periodic boundary conditions on $\psi_{\pm}$. 
While the system remains gapless under the anti-periodic boundary conditions, the vacuum is unique.

On the other hand, if $m\ne 0$, the mass term explicitly breaks the fermionic subsystem symmetry. The system becomes trivially gapped in the absence of boundaries, as we will demonstrate below.

Since the massive $\psi$ theory \eqref{eq:action} is a free field theory, we can easily solve it by the mode-expansion.  The expressions of $\psi_{\pm}$ in terms of eigen-modes are given by
\begin{align}
	\psi_+=&\sum_{\vec{k}} \sqrt{\frac{1}{2V}\frac{\omega_{\vec{k}}+k_xk_yk_z}{\omega_{\vec{k}}}}\left(a_{\vec{k}}\exp(i\vec{k}\cdot\vec{x}-i\omega_{\vec{k}} t)+a_{\vec{k}}^\dag\exp(-i\vec{k}\cdot\vec{x}+i\omega_{\vec{k}} t)\right),\qquad\qquad\qquad\nonumber\\
	\psi_-=&i\sign m \sum_{\vec{k}}\sqrt{\frac{1}{2V}\frac{\omega_{\vec{k}}-k_xk_yk_z}{\omega_{\vec{k}}}}\left(a_{\vec{k}}\exp(i\vec{k}\cdot\vec{x}-i\omega_{\vec{k}} t)-a_{\vec{k}}^\dag\exp(-i\vec{k}\cdot\vec{x}+i\omega_{\vec{k}} t)\right),\notag\\
	 V:=& l_xl_yl_z,
	\label{eq:noboundary_psi_mode}
\end{align}
where the wave numbers $\vec{k}=(k_x,k_y,k_z)$ take the following values due to the anti-periodic boundary condition:
\begin{align}
	k_i=\frac{2\pi}{l_i}\left(n_i+\frac12\right), \qquad n_i\in\mathbb{Z},\qquad i=x,y,z.
\end{align}
The dispersion relation is given by
\begin{align}
	\omega_{\vec{k}}=\sqrt{k_x^2k_y^2k_z^2+m^2}.
	\label{eq:omegak}
\end{align}

The anti-commutation relations of the mode operators $a_{\vec{k}},a_{\vec{k}}^{\dag}$ are given by
\begin{align}
	\left\{a_{\vec{k}},a_{\vec{k'}}^\dag\right\}=\delta_{\vec{k},\vec{k'}}.
\end{align}
The Hamiltonian of this system is given by
\begin{align}
	H=\int d^3x\left(\frac{1}{2}i\psi_+\partial_x\partial_y\partial_z\psi_+-\frac{1}{2}i\psi_-\partial_x\partial_y\partial_z\psi_-+im\psi_-\psi_+\right)
	=\sum_{\vec{k}}\omega_{\vec{k}}a_{\vec{k}}^{\dag} a_{\vec{k}}.
\end{align}

From dispersion relation \eqref{eq:omegak}, we find
\begin{align}
  \omega_{\vec{k}}\geq |m|,
\end{align}
indicating that the system is gapped. The ground state is annihilated by $a_{\vec{k}}$ for all $\vec{k}$ and it is unique.  Thus, the system is trivially gapped in the absence of boundaries.

\section{The system with boundary}
\label{sec:boundary}

In this section, we examine the effects of introducing boundaries to the system. First, we explore the possible boundary conditions. Subsequently, we analyze the spectrum of the system in the presence of boundaries.

\subsection{Boundary conditions}

The space we consider here is $I \times T^2$.  Here, $I$ is an interval defined by $0\leq x\leq l_x$ with boundaries at $x=0$ and $x=l_x$.  The coordinates of the torus $T^2$, denoted by $y,z$, are identified as $y\sim y+l_y$ and $z\sim z+l_z$.

The action of the massive $\psi$ theory is given by
\begin{equation}
	S=\int d^3xdt( i\psi_+\partial_- \psi_++i\psi_-\partial_+\psi_--im\psi_-\psi_+).
	\label{eq:boundaryaction}
\end{equation}
For simplicity, we impose anti-periodic boundary conditions on $\psi_{\pm}$ along the $y$ and $z$ directions:
\begin{align}
	\psi_{\pm}(x,y,z)=-\psi_{\pm}(x,y+l_y,z),\nonumber\\
	\psi_{\pm}(x,y,z)=-\psi_{\pm}(x,y,z+l_z).
	\label{eq:antiperiodic-yz}
\end{align}

To determine the possible boundary conditions at $x=0$ and $x=l_x$, we apply the variational principle. 
The variation of the action \eqref{eq:boundaryaction} is given by
\begin{align}
  \delta S=&\int d^3xdt\left[\ 2i\delta\psi_+\partial_-\psi_++2i\delta\psi_-\partial_+\psi_--im\delta\psi_-\psi_+-im\psi_-\delta\psi_+\right]\nonumber\\
  +&\int dydzdt[\delta\psi_+\partial_y\partial_z\psi_+-\delta\psi_-\partial_y\partial_z\psi_-]^{x=l_x}_{x=0}.
  \label{eq:boundary_diff}
\end{align}
The boundary conditions are chosen to ensure that the boundary term in \eqref{eq:boundary_diff} vanishes:
\begin{align}
	\int dydzdt[\delta\psi_+\partial_y\partial_z\psi_+-\delta\psi_-\partial_y\partial_z\psi_-]^{x=l_x}_{x=0}.
	\label{eq:boundarycondrq}
\end{align}
From this, we identify the possible boundary conditions as:
\begin{align}
	\partial_y\partial_z\psi_+=\pm\partial_y\partial_z\psi_-|_{x=0,l_x}.
	\label{eq:boundary_condition}
\end{align}
Actually, the boundary conditions \eqref{eq:boundary_condition} imply
\begin{align}
	&\int dydz\left[\delta\psi_+\partial_y\partial_z\psi_+-\delta\psi_-\partial_y\partial_z\psi_-\right]_{x=0}=
	\int dydz\left[(\delta\psi_+ \mp \delta\psi_-)\partial_y\partial_z\psi_+\right]_{x=0}\notag\\
	&\qquad =
	\int dydz\left[\partial_y\partial_z(\delta\psi_+ \mp \delta\psi_-)\psi_+\right]_{x=0}=0.
\end{align}
The same holds at $x=l_x$.  Therefore, the boundary conditions \eqref{eq:boundary_condition} satisfy \eqref{eq:boundarycondrq} and the variational principle is satisfied.

There are four possible boundary conditions based on the signs in \eqref{eq:boundary_condition}:
\begin{enumerate}
  \renewcommand{\theenumi}{(\roman{enumi})}
  \item \label{bc1}$\partial_y\partial_z\psi_+=\partial_y\partial_z\psi_-|_{x=0,l_x}.$
  \item \label{bc2} $\partial_y\partial_z\psi_+=\partial_y\partial_z\psi_-|_{x=0},\quad\partial_y\partial_z\psi_+=-\partial_y\partial_z\psi_-|_{x=l_x}.$
  \item \label{bc3} $\partial_y\partial_z\psi_+=-\partial_y\partial_z\psi_-|_{x=0,l_x}.$
  \item \label{bc4} $\partial_y\partial_z\psi_+=-\partial_y\partial_z\psi_-|_{x=0},\quad\partial_y\partial_z\psi_+=+\partial_y\partial_z\psi_-|_{x=l_x}.$
\end{enumerate}
However, the boundary conditions \ref{bc3} and \ref{bc4} are derived by substituting $m\rightarrow-m,\psi_{\pm}\rightarrow \pm\psi_{\pm}$ to the boundary conditions \ref{bc1} and \ref{bc2}, respectively.  Thus, we focus on the boundary conditions \ref{bc1} and \ref{bc2} in the following.

\subsection{Spectrum of the system with boundary}

Let us find the spectrum of the system with boundary. The bulk theory is a free theory and the boundary conditions \ref{bc1} and \ref{bc2} are linear. Therefore, we can solve the system by the mode expansion.  In order to find the spectrum, it is enough to find the eigen-modes and the dispersion relation.

We now examine the eigen-modes of the system in detail with the boundary conditions \ref{bc1} and \ref{bc2}.

\subsubsection{\texorpdfstring{Case \ref{bc1}}{ Case (i)}}
\label{section++}

Here, we analyze the system with the boundary condition \ref{bc1}.  
There are two classes of mode functions to consider: the bulk modes and the boundary-localized modes.
Both of them satisfy the equations of motion derived from \eqref{eq:boundary_diff}:
\begin{align}
	2\partial_-\psi_++m\psi_-=0,\nonumber\\
	2\partial_+\psi_--m\psi_+=0.
	\label{eq:boundary_eom}
\end{align}

Let us first consider the bulk modes $\psibulk_{\pm}$.  
A plane wave solution of the equations of motion \eqref{eq:boundary_eom} is given by
\begin{align}
	\psibulk_{+}&=A\exp(i\vec{k}\cdot\vec{x}-i\omega t),\nonumber\\
	\psibulk_{-}&=\frac{1}{m}(i\omega-ik_xk_yk_z)A\exp(i\vec{k}\cdot\vec{x}-i\omega t),\nonumber\\
	\omega&=\pm \omega_{\vec{k}},\quad \omega_{\vec{k}}:=\sqrt{k_x^2k_y^2k_z^2+m^2},
\end{align}
where $A$ is a constant and $\vec{k}=(k_x,k_y,k_z)$ represents the wave vector.

To satisfy the boundary condition, we make an ansatz that the bulk modes are given by a superposition of incident and reflected waves:
\begin{align}
	\psibulk_{+}&=A\exp(i\vec{k}\cdot\vec{x}-i\omega t)+B\exp(-ik_xx+ik_y+ik_zz-i\omega t),\nonumber\\
	\psibulk_{-}&=\frac{1}{m}(i\omega-ik_xk_yk_z)A\exp(i\vec{k}\cdot\vec{x}-i\omega t)+\frac{1}{m}(i\omega+ik_xk_yk_z)B\exp(-ik_xx+ik_yy+ik_zz-i\omega t),
	\label{eq:bulk_eom}
\end{align}
where $A$ and $B$ are constants.

We substitute this ansatz for the boundary condition \ref{bc1} and obtain 
\begin{align}
	&(m- i\omega+i k_xk_yk_z)A+(m- i\omega - ik_xk_yk_z)B=0,\notag\\
	&(m- i\omega+i k_xk_yk_z)\exp(ik_x l_x)A+(m - i\omega - ik_xk_yk_z)\exp(-ik_x l_x)B=0.
	\label{eq:boundary_x=0bulkbou}
\end{align}
In order for the above equations to have non-trivial solutions, the determinant of the coefficient matrix must vanish.
\begin{align}
	\det\begin{pmatrix}
	m- i\omega+i k_xk_yk_z & m- i\omega- ik_xk_yk_z\\
	(m- i\omega+i k_xk_yk_z)\exp(ik_x l_x) & (m-i\omega- ik_xk_yk_z)\exp(-ik_x l_x)\\
	\end{pmatrix}
	=0.
\end{align}
This equation is equivalent to
\begin{align}
	(m-i\omega+ik_xk_yk_z)(m-i\omega-ik_xk_yk_z)\left(\exp(ik_x l_x)-\exp(-ik_x l_x)\right)=0.
\end{align}
Since $m\ne 0$ and $\omega, k_x, k_y, k_z$ are real, the above equation is equivalent to
\begin{align}
	\exp(ik_x l_x)-\exp(-ik_xx)=0.	
\end{align}
From this condition, the possible values of $k_x$ are given by
\begin{align}
	k_x=\frac{n_x\pi}{l_x}\ \ \ ,n_x\in\mathbb{Z}.
\end{align}
Actually, $n_x=n$ and $n_x=-n$ give the same mode function as seen in Eq.~\eqref{eq:bulk_eom}.  Thus, we choose $n_x>0$ as the representative.  Moreover, $n_x=0$ is not valid since the mode function \eqref{eq:bulk_eom} vanishes in this case. Therefore, we obtain
\begin{align}
	\omega&=\pm\sqrt{k_x^2k_y^2k_z^2+m^2},\notag\\
	k_x=\frac{n_x\pi}{l_x},\quad k_y&=\frac{(2n_y+1)\pi}{l_y},\quad k_z=\frac{(2n_z+1)\pi}{l_z},\qquad n_x,n_y,n_z\in\mathbb{Z}\ \ ,n_x > 0,
	\label{eq:boundary_ibulk}
\end{align}
for the spectrum corresponding to the bulk modes.

Next, let us consider the boundary-localized modes, denoted by $\psil_{\pm}$.  We propose the following ansatz for these modes:
\begin{align}
  \psil_{+}&=A\exp(-\Lambda x+ik_y+ik_zz-i\omega t)+B\exp(\Lambda x+ik_y+ik_zz-i\omega t)\nonumber\\
  \psil_{-}&=\frac{1}{m}(i\omega+\Lambda k_yk_z)A\exp(-\Lambda x+ik_y+ik_zz-i\omega t)\nonumber\\
  &\ \ \ \ \ \ \ \ \ \ +\frac{1}{m}(i\omega -\Lambda k_yk_z)B\exp(\Lambda x+ik_y+ik_zz-i\omega t),\nonumber\\
  &\omega=\pm\omegal(\Lambda,k_y,k_z),\quad \omegal(\Lambda,k_y,k_z) := \sqrt{-\Lambda^2k_y^2k_z^2+m^2},
  \label{eq:local_eom}
\end{align}
where $A,B$ are constants and $\Lambda$ is a positive real number.

The boundary condition \ref{bc1} leads to the following relations:
\begin{align}
&(m- i\omega-\Lambda k_yk_z)A + (m- i\omega+\Lambda k_yk_z)B=0,\notag\\
  &(m- i\omega-\Lambda k_yk_z)\exp(-\Lambda l_x)A+ (m-i\omega+\Lambda k_yk_z)\exp(\Lambda l_x)B=0.
  \label{eq:+-boundary}
\end{align}
For non-trivial solutions to exist, the determinant of the coefficient matrix in \eqref{eq:+-boundary} must vanish.  This condition is equivalent to:
\begin{align}
	(m-i\omega+\Lambda k_yk_z)(m-i\omega-\Lambda k_yk_z)(\exp(\Lambda l_x)-\exp(-\Lambda l_x))=0.
\end{align}
The solutions of this equation are given by  $\Lambda=0,\pm\frac{m}{k_yk_z}$.  
\begin{itemize}
	\item First, the solution $\Lambda=0$ is invalid, as the mode function vanishes in this case.
	\item Second, For $\Lambda=\frac{m}{k_yk_z}$, we find $B=0$ and $\omega=0$ from \eqref{eq:+-boundary} and \eqref{eq:local_eom}, respectively. Since $\Lambda>0$, this solution is valid only when $\frac{m}{k_yk_z}>0$. The corresponding mode function is:
	\begin{align}
		\psil_{+}&=A\exp\left(-\frac{m}{k_yk_z}x+ik_y+ik_zz\right),\nonumber\\
		\psil_{-}&=A\exp\left(-\frac{m}{k_yk_z}x+ik_yy+ik_zz\right).
		\label{edgemodeR0}
	\end{align}
	This solution is localized at $x=0$.
	\item Third, For $\Lambda=-\frac{m}{k_yk_z}$, we find $A=0$ and $\omega=0$ from \eqref{eq:+-boundary} and \eqref{eq:local_eom}, respectively. Since $\Lambda>0$, this solution is valid only when $\frac{m}{k_yk_z}<0$. The corresponding mode function is:
	\begin{align}
		\psil_{+}&=B\exp\left(-\frac{m}{k_yk_z}x+ik_y+ik_zz\right),\nonumber\\
		\psil_{-}&=B\exp\left(-\frac{m}{k_yk_z}x+ik_yy+ik_zz\right).
		\label{edgemodeRl}
	\end{align}	
	This solution is localized at $x=l_x$.
\end{itemize}
The frequencies of these modes are exactly zero.

It follows that the system with this boundary exhibits exponentially large vacuum degeneracy. As demonstrated above, there is one Majorana zero mode for each pair $(k_y,k_z)$. Let $N$ be the number of such pairs, i.e., the number of Majorana zero modes. This number scales as $N\sim\frac{l_yl_z}{a^2}$ with the lattice spacing $a$. The vacuum degeneracy is given by $2^{\frac{1}{2}N}$. 
Consequently, the ground states of the system with this boundary are exponentially degenerate. 
Furthermore, the residual entropy is proportional to the area of the boundary.

Recall that the analogous boundary condition of the $(1+1)$-dimensional  Majorana fermion is forbidden due to the existence of an odd number of Majorana zero modes.
However, the boundary condition in our theory is not forbidden since the number of the Majorana zero modes is even, as shown below. 
Actually, the complex conjugates of the mode functions \eqref{edgemodeR0} and \eqref{edgemodeRl} for $(k_y,k_z)$ are identical to the mode functions for $(-k_y,-k_z)$. 
Therefore, precisely speaking, there is a Dirac zero mode or two Majorana zero modes for a pair $(k_y,k_z)$ and $(-k_y,-k_z)$ if $(k_y,k_z)\ne (0,0)$.
Additionally, note that $(k_y,k_z)\ne (0,0)$ for our anti-periodic boundary condition \eqref{eq:antiperiodic-yz}.
Consequently, the number of Majorana zero modes are even.

\subsubsection{\texorpdfstring{Case \ref{bc2}}{ Case (ii)}}

Here, we analyze the system with boundary condition \ref{bc2}.

The ansatz for the bulk modes remains the same as in the previous case \eqref{eq:bulk_eom}. The boundary condition \ref{bc2} leads to the following relations:
\begin{align}
  	&(m- i\omega+i k_xk_yk_z)A+(m- i\omega - ik_xk_yk_z)B=0,\notag\\
	&(m+ i\omega-i k_xk_yk_z)\exp(ik_x l_x)A + (m + i\omega + ik_xk_yk_z)\exp(-ik_x l_x)B=0.
	\label{eq:boundary(ii)}
\end{align}
In order for the above equations to have non-trivial solutions, the determinant of the coefficient matrix must vanish:
\begin{align}
  \det\begin{pmatrix}
  m- i\omega+i k_xk_yk_z & m- i\omega- ik_xk_yk_z\\
  (m+i\omega-i k_xk_yk_z)\exp(ik_x l_x) & (m+i\omega+ik_xk_yk_z)\exp(-ik_x l_x)\\
  \end{pmatrix}
  =0.
\end{align}
This condition simplifies to:
\begin{align}
  \tan k_xl_x = \frac{k_xk_yk_z}{m}.
  \label{eq:+-bulk_kxsol}
\end{align}
The general solution of this equation cannot be expressed in terms of elementary functions. Approximated solutions are available in the limiting cases:
\begin{align}
	k_x &\approx \frac{n\pi}{l_x},
	\qquad &\text{if} \qquad |k_x|\ll\left|\frac{m}{k_yk_z}\right|,  \notag\\
	k_x &\approx \frac{(2n+1)\pi}{2l_x},
	\qquad &\text{if} \qquad 
	|k_x|\gg\left|\frac{m}{k_yk_z}\right|,\\
	& &n\in \Zb.  \notag
\end{align}
Although $k_x=0$ is an exact solution of \eqref{eq:+-bulk_kxsol}, the corresponding mode function is $\psil_\pm=0$, and therefore not suitable as a valid mode function.

The ansatz for the boundary-localized modes is the same as the previous case \eqref{eq:local_eom}. The boundary condition \ref{bc2} leads to the following relations:
\begin{align}
  &(m- i\omega-\Lambda k_yk_z)A + (m- i\omega+\Lambda k_yk_z)B=0,\notag\\
  &(m + i\omega + \Lambda k_yk_z)\exp(-\Lambda l_x)A+ (m+i\omega-\Lambda k_yk_z)\exp(\Lambda l_x)B=0.	
\end{align}
For non-trivial solutions to exist, the determinant of the coefficient matrix must vanish:
\begin{align}
  \det\begin{pmatrix}
  m- i\omega-\Lambda k_yk_z & m- i\omega +\Lambda k_yk_z\\
  (m+ i\omega+\Lambda k_yk_z)\exp(-\Lambda l_x) & (m+i\omega-\Lambda k_yk_z)\exp(\Lambda l_x)\\
  \end{pmatrix}
  =0.
\end{align}
This condition simplifies to:
\begin{align}
	\tanh \Lambda l_x = \frac{\Lambda k_y k_z}{m}.
  \label{eq:boundary_spect}
\end{align}
$\Lambda=0$ is an exact solution of \eqref{eq:boundary_spect}, but the corresponding mode function for this solution is $\psil_\pm=0$, and not suitable as a mode function.
A non-zero solution exists for Eq.~\eqref{eq:boundary_spect} if and only if $1 < \frac{ml_x}{k_y k_z}$.
Assuming that the system size is large enough to satisfy $1\ll\frac{ml_x}{k_y k_z}$,\footnote{We implicitly assume the existence of a cutoff, ensuring an upper bound for $|k_y k_z|$. Thus, this condition is satisfied if $l_x$ is chosen sufficiently large.} we can evaluate the frequency of the boundary-localized mode.  The solution of Eq.~\eqref{eq:boundary_spect} is expressed as
\begin{align}
	\Lambda = \Lambda_0 + O(e^{-2\Lambda_0 l_x}),\quad
	\Lambda_0 := \frac{m}{k_y k_z}.
\end{align}
Note that the reminder term $O(e^{-2\Lambda_0 l_x})$ in the expression of $\Lambda$ is negative, leading to $\Lambda < \Lambda_0$.
The frequency of the boundary-localized mode is given by
\begin{align}
	\omega = \pm \omegal(\Lambda,k_y,k_z),\quad
	\omegal(\Lambda,k_y,k_z) := \sqrt{-\Lambda^2k_y^2k_z^2+m^2}= O(e^{-\Lambda_0 l_x}).
\end{align}
Since $\omegal(\Lambda,k_y,k_z)>0$, the boundary-localized modes have exponentially small positive and negative frequencies.

The system with this boundary also exhibits exponentially large vacuum degeneracy.
As shown above, there are two approximate zero modes for a pair $(k_y,k_z)$ that satisfy $\frac{k_yk_z}{m}>0$ and no approximate zero modes for $(k_y,k_z)$ that satisfy $\frac{k_yk_z}{m}<0$.  
The total number of approximate zero modes are the same as in Case \ref{bc1}.  
Consequently, the ground state degeneracy is also the same as that in Case \ref{bc1}.

This system is similar to the $(1+1)$-dimensional Majorana fermion with the Neveu-Schwarz (NS) boundary condition. In the $(1+1)$-dimensional case, the number of Majorana zero modes depends on the sign of the mass term and can be either zero or two.
In contrast, for the present system, a large number of Majorana zero modes exist regardless of the sign of $m$.

\section{Conclusion and discussion}
\label{sec:conclusion}

In this paper, we investigated the massive $\psi$ theory \eqref{eq:action} with boundaries.
While the system is trivially gapped in the absence of boundaries, we explored the effects of introducing boundary conditions \ref{bc1} and \ref{bc2}.
For Case \ref{bc1}, there are a large number of exact zero modes, resulting a huge vacuum degeneracy.
The residual entropy is proportional to the area of the boundary.
Non-zero modes exhibit a gap of $|m|$, ensuring that the system remains gapped.
For Case \ref{bc2}, although the zero modes are not exact but approximate, the vacuum degeneracy and the gap are the same as in Case \ref{bc1}.

This behavior is reminiscent of symmetry protected topological (SPT) phases, where the system is trivially gapped in the absence of boundaries but becomes gapless or exhibits vacuum degeneracy in the presence of boundaries.  One may suspect that the massive $\psi$ theory belongs to an SPT phase or an SSPT phase.  However, there are notable differences:
\begin{itemize}
	\item Our system on the space with boundary has a vacuum degeneracy, regardless of the sign of $m$.  In contrast, typical SPT phases become trivial phase if the sign of the mass is reversed.
	\item The symmetry that protects the vacuum degeneracy of the massive $\psi$ theory is not yet identified.  It is not the fermionic subsystem symmetry, as it is explicitly broken by the mass term.
	\item No inconsistencies were found in the boundary condition \ref{bc1}.  This contrasts with the analogous boundary condition for the $(1+1)$-dimensional Majorana fermion, which is inconsistent due to the existence of odd number of Majorana zero modes.
 \end{itemize}
Identifying the symmetry that protects the vacuum degeneracy of the massive $\psi$ theory is an intriguing direction for future research.  Possible candidates include discrete symmetries such as discrete translations or 90-degree rotations.

In this paper, we consider only the boundary orthogonal to the $x$-direction.
It would be interesting to introduce the boundaries orthogonal to the $y$ or $z$ direction and investigate the behavior of the system at corners.
Furthermore, studying oblique boundaries is also promising direction, as our system does not have the continuous rotational symmetry.	
These questions are left for future work.

\subsection*{Acknowledgement}
We would like to thank Hiromi Ebisu, Ken Shiozaki for useful discussions.
The work of S.Y. was supported in part by JSPS KAKENHI Grant Number 21K03574.

\bibliographystyle{utphys}
\bibliography{ref}

\providecommand{\href}[2]{#2}\begingroup\raggedright\begin{thebibliography}{10}

\bibitem{Chamon:2004lew}
C.~Chamon, ``{Quantum Glassiness},'' \href{https://dx.doi.org/10.1103/physrevlett.94.040402}{{\em Phys. Rev. Lett.} {\bfseries 94} no.~4, (2005) 040402}, \href{https://arxiv.org/abs/cond-mat/0404182}{{\ttfamily arXiv:cond-mat/0404182}}.

\bibitem{Haah:2011drr}
J.~Haah, ``{Local stabilizer codes in three dimensions without string logical operators},'' \href{https://dx.doi.org/10.1103/physreva.83.042330}{{\em Phys. Rev. A} {\bfseries 83} no.~4, (2011) 042330}, \href{https://arxiv.org/abs/1101.1962}{{\ttfamily arXiv:1101.1962 [quant-ph]}}.

\bibitem{Nandkishore:2018sel}
R.~M. Nandkishore and M.~Hermele, ``{Fractons},'' \href{https://dx.doi.org/10.1146/annurev-conmatphys-031218-013604}{{\em Ann. Rev. Condensed Matter Phys.} {\bfseries 10} (2019) 295--313}, \href{https://arxiv.org/abs/1803.11196}{{\ttfamily arXiv:1803.11196 [cond-mat.str-el]}}.

\bibitem{Pretko:2020cko}
M.~Pretko, X.~Chen, and Y.~You, ``{Fracton Phases of Matter},'' \href{https://dx.doi.org/10.1142/S0217751X20300033}{{\em Int. J. Mod. Phys. A} {\bfseries 35} no.~06, (2020) 2030003}, \href{https://arxiv.org/abs/2001.01722}{{\ttfamily arXiv:2001.01722 [cond-mat.str-el]}}.

\bibitem{Gromov:2022cxa}
A.~Gromov and L.~Radzihovsky, ``{Colloquium: Fracton matter},'' \href{https://dx.doi.org/10.1103/RevModPhys.96.011001}{{\em Rev. Mod. Phys.} {\bfseries 96} no.~1, (2024) 011001}, \href{https://arxiv.org/abs/2211.05130}{{\ttfamily arXiv:2211.05130 [cond-mat.str-el]}}.

\bibitem{Yamaguchi:2021qrx}
S.~Yamaguchi, ``{Supersymmetric quantum field theory with exotic symmetry in 3+1 dimensions and fermionic fracton phases},'' \href{https://dx.doi.org/10.1093/ptep/ptab037}{{\em PTEP} {\bfseries 2021} no.~6, (2021) 063B04}, \href{https://arxiv.org/abs/2102.04768}{{\ttfamily arXiv:2102.04768 [hep-th]}}.

\bibitem{Seiberg:2019vrp}
N.~Seiberg, ``{Field Theories With a Vector Global Symmetry},'' \href{https://dx.doi.org/10.21468/SciPostPhys.8.4.050}{{\em SciPost Phys.} {\bfseries 8} no.~4, (2020) 050}, \href{https://arxiv.org/abs/1909.10544}{{\ttfamily arXiv:1909.10544 [cond-mat.str-el]}}.

\bibitem{Seiberg:2020bhn}
N.~Seiberg and S.-H. Shao, ``{Exotic Symmetries, Duality, and Fractons in 2+1-Dimensional Quantum Field Theory},'' \href{https://arxiv.org/abs/2003.10466}{{\ttfamily arXiv:2003.10466 [cond-mat.str-el]}}.

\bibitem{Seiberg:2020wsg}
N.~Seiberg and S.-H. Shao, ``{Exotic $U(1)$ Symmetries, Duality, and Fractons in 3+1-Dimensional Quantum Field Theory},'' \href{https://dx.doi.org/10.21468/SciPostPhys.9.4.046}{{\em SciPost Phys.} {\bfseries 9} no.~4, (2020) 046}, \href{https://arxiv.org/abs/2004.00015}{{\ttfamily arXiv:2004.00015 [cond-mat.str-el]}}.

\bibitem{Gorantla:2020xap}
P.~Gorantla, H.~T. Lam, N.~Seiberg, and S.-H. Shao, ``{More Exotic Field Theories in 3+1 Dimensions},'' \href{https://dx.doi.org/10.21468/SciPostPhys.9.5.073}{{\em SciPost Phys.} {\bfseries 9} (2020) 073}, \href{https://arxiv.org/abs/2007.04904}{{\ttfamily arXiv:2007.04904 [cond-mat.str-el]}}.

\bibitem{Katsura:2022xkg}
H.~Katsura and Y.~Nakayama, ``{Spontaneously broken supersymmetric fracton phases with fermionic subsystem symmetries},'' \href{https://dx.doi.org/10.1007/JHEP08(2022)072}{{\em JHEP} {\bfseries 08} (2022) 072}, \href{https://arxiv.org/abs/2204.01924}{{\ttfamily arXiv:2204.01924 [hep-th]}}.

\bibitem{Honda:2022shd}
M.~Honda and T.~Nakanishi, ``{Scalar, fermionic and supersymmetric field theories with subsystem symmetries in d + 1 dimensions},'' \href{https://dx.doi.org/10.1007/JHEP03(2023)188}{{\em JHEP} {\bfseries 03} (2023) 188}, \href{https://arxiv.org/abs/2212.13006}{{\ttfamily arXiv:2212.13006 [hep-th]}}.

\bibitem{you2019building}
Y.~You and F.~von Oppen, ``Building fracton phases by majorana manipulation,'' \href{https://dx.doi.org/10.1103/PhysRevResearch.1.013011}{{\em Physical Review Research} {\bfseries 1} no.~1, (2019) 013011}.

\bibitem{Tantivasadakarn:2020lhq}
N.~Tantivasadakarn, ``{Jordan-Wigner Dualities for Translation-Invariant Hamiltonians in Any Dimension: Emergent Fermions in Fracton Topological Order},'' \href{https://dx.doi.org/10.1103/PhysRevResearch.2.023353}{{\em Phys. Rev. Res.} {\bfseries 2} no.~2, (2020) 023353}, \href{https://arxiv.org/abs/2002.11345}{{\ttfamily arXiv:2002.11345 [cond-mat.str-el]}}.

\bibitem{Shirley:2020ass}
W.~Shirley, ``{Fractonic order and emergent fermionic gauge theory},'' \href{https://arxiv.org/abs/2002.12026}{{\ttfamily arXiv:2002.12026 [cond-mat.str-el]}}.

\bibitem{Cao:2022lig}
W.~Cao, M.~Yamazaki, and Y.~Zheng, ``{Boson-fermion duality with subsystem symmetry},'' \href{https://dx.doi.org/10.1103/PhysRevB.106.075150}{{\em Phys. Rev. B} {\bfseries 106} no.~7, (2022) 075150}, \href{https://arxiv.org/abs/2206.02727}{{\ttfamily arXiv:2206.02727 [cond-mat.str-el]}}.

\bibitem{Callan:1984sa}
C.~G. Callan, Jr. and J.~A. Harvey, ``{Anomalies and Fermion Zero Modes on Strings and Domain Walls},'' \href{https://dx.doi.org/10.1016/0550-3213(85)90489-4}{{\em Nucl. Phys. B} {\bfseries 250} (1985) 427--436}.

\bibitem{You:2018oai}
Y.~You, T.~Devakul, F.~J. Burnell, and S.~L. Sondhi, ``{Subsystem symmetry protected topological order},'' \href{https://dx.doi.org/10.1103/PhysRevB.98.035112}{{\em Phys. Rev. B} {\bfseries 98} no.~3, (2018) 035112}, \href{https://arxiv.org/abs/1803.02369}{{\ttfamily arXiv:1803.02369 [cond-mat.str-el]}}.

\bibitem{Devakul:2018fhz}
T.~Devakul, D.~J. Williamson, and Y.~You, ``{Classification of subsystem symmetry-protected topological phases},'' \href{https://dx.doi.org/10.1103/PhysRevB.98.235121}{{\em Phys. Rev. B} {\bfseries 98} no.~23, (2018) 235121}, \href{https://arxiv.org/abs/1808.05300}{{\ttfamily arXiv:1808.05300 [cond-mat.str-el]}}.

\bibitem{Devakul:2019duj}
T.~Devakul, W.~Shirley, and J.~Wang, ``{Strong planar subsystem symmetry-protected topological phases and their dual fracton orders},'' \href{https://dx.doi.org/10.1103/PhysRevResearch.2.012059}{{\em Phys. Rev. Res.} {\bfseries 2} no.~1, (2020) 012059}, \href{https://arxiv.org/abs/1910.01630}{{\ttfamily arXiv:1910.01630 [cond-mat.str-el]}}.

\bibitem{Burnell:2021reh}
F.~J. Burnell, T.~Devakul, P.~Gorantla, H.~T. Lam, and S.-H. Shao, ``{Anomaly inflow for subsystem symmetries},'' \href{https://dx.doi.org/10.1103/PhysRevB.106.085113}{{\em Phys. Rev. B} {\bfseries 106} no.~8, (2022) 085113}, \href{https://arxiv.org/abs/2110.09529}{{\ttfamily arXiv:2110.09529 [cond-mat.str-el]}}.

\bibitem{Yamaguchi:2021xeq}
S.~Yamaguchi, ``{Gapless edge modes in (4+1)-dimensional topologically massive tensor gauge theory and anomaly inflow for subsystem symmetry},'' \href{https://dx.doi.org/10.1093/ptep/ptac032}{{\em PTEP} {\bfseries 2022} no.~3, (2022) 033B08}, \href{https://arxiv.org/abs/2110.12861}{{\ttfamily arXiv:2110.12861 [hep-th]}}.

\bibitem{Shimamura:2024kwf}
S.~Shimamura, ``{Anomaly of subsystem symmetries in exotic and foliated BF theories},'' \href{https://dx.doi.org/10.1007/JHEP06(2024)002}{{\em JHEP} {\bfseries 06} (2024) 002}, \href{https://arxiv.org/abs/2404.10601}{{\ttfamily arXiv:2404.10601 [cond-mat.str-el]}}.

\bibitem{Okuda:2024azp}
T.~Okuda, A.~Parayil~Mana, and H.~Sukeno, ``{Anomaly inflow for CSS and fractonic lattice models and dualities via cluster state measurement},'' \href{https://dx.doi.org/10.21468/SciPostPhys.17.4.113}{{\em SciPost Phys.} {\bfseries 17} no.~4, (2024) 113}, \href{https://arxiv.org/abs/2405.15853}{{\ttfamily arXiv:2405.15853 [quant-ph]}}.

\bibitem{Ebisu:2024eew}
H.~Ebisu, M.~Honda, and T.~Nakanishi, ``{Anomaly inflow for dipole symmetry and higher form foliated field theories},'' \href{https://dx.doi.org/10.1007/JHEP09(2024)061}{{\em JHEP} {\bfseries 09} (2024) 061}, \href{https://arxiv.org/abs/2406.04919}{{\ttfamily arXiv:2406.04919 [cond-mat.str-el]}}.

\bibitem{Kitaev:2000nmw}
A.~Kitaev, ``{Unpaired Majorana fermions in quantum wires},'' \href{https://dx.doi.org/10.1070/1063-7869/44/10S/S29}{{\em Phys. Usp.} {\bfseries 44} no.~10S, (2001) 131--136}, \href{https://arxiv.org/abs/cond-mat/0010440}{{\ttfamily arXiv:cond-mat/0010440}}.

\end{thebibliography}\endgroup
\end{document}